\documentclass[aps,amsmath,twocolumn,amssymb,floatfixng,showpacs,
superscriptaddress,footinbib]{revtex4-1}
\pdfoutput=1
\usepackage[dvips]{graphics}
\usepackage{bm}
\usepackage{float}
\usepackage{epsfig}
\usepackage{enumerate}
\usepackage{subfigure}
\usepackage{amsmath}
\usepackage{color}
\usepackage{braket}
\usepackage{graphicx}
\usepackage{float}
\usepackage[colorlinks=true,linktoc=page,linkcolor=red,citecolor=blue,urlcolor=magenta]{hyperref}

\newcommand{\vect}[1]{\bm{\mathrm{#1}}}

\newcommand{\ie}{{\it i.e.,\,\,}}
\newcommand{\eg}{{\it e.g.,~}}

\newcommand\bea{\begin{eqnarray}}
\newcommand\eea{\end{eqnarray}}
\newcommand\beq{\begin{equation}}  
\newcommand\eeq{\end{equation}}

\newcommand{\non}{\nonumber}

\begin{document} 

\title{Floquet generation of Second Order Topological  Superconductor} 
 \author{Arnob Kumar Ghosh}
\email{arnob@iopb.res.in}
\affiliation{Institute of Physics, Sachivalaya Marg, Bhubaneswar-751005, India}
\affiliation{Homi Bhabha National Institute, Training School Complex, Anushakti Nagar, Mumbai 400094, India}
\author{Tanay Nag}
\email{tnag@sissa.it}
\affiliation{SISSA, via Bonomea 265, 34136 Trieste, Italy}
\author{Arijit Saha}
\email{arijit@iopb.res.in}
\affiliation{Institute of Physics, Sachivalaya Marg, Bhubaneswar-751005, India}
\affiliation{Homi Bhabha National Institute, Training School Complex, Anushakti Nagar, Mumbai 400094, India}

\begin{abstract}
We theoretically investigate the Floquet generation of second-order topological superconducting (SOTSC) phase, hosting Majorana corner modes (MCMs), considering a quantum spin Hall insulator (QSHI) with proximity induced superconducting $s$-wave pairing in it. Our dynamical prescription consists of the periodic kick in time-reversal symmetry breaking in-plane magnetic field and four-fold rotational symmetry breaking mass term in the bulk while these Floquet MCMs are preserved by anti-unitary particle-hole symmetry. The first driving protocol always leads to four zero energy MCMs (\ie one Majorana state per corner) as a sign of a {\it{strong}} SOTSC phase. Interestingly, the second protocol can result in a {\it{weak}} SOTSC phase, harbouring eight zero energy MCMs (two Majorana states per corner), in addition to the {\it{strong}} SOTSC phase. We characterize the topological nature of these phases by Floquet quadrupolar moment and Floquet Wannier spectrum. We believe that relying on the recent experimental advancement in the driven systems and proximity induced superconductivity, our schemes may be possible to test in the future. 
\end{abstract}

\maketitle


\section{Introduction}{\label{sec:I}}

In recent times, the topological superconductors (TSC), hosting Majorana zero-energy modes at their boundaries, have attracted enormous attention both theoretically and experimentally due to their connection with non-Abelian exchange statistics and potential applications in topological quantum computation~\cite{Kitaev_2001,qi2011topological,hasan2010colloquium,nayak08,das2012zero,Deng1557}. The heterostructures of materials with strong spin-orbit coupling such as topological insulator, semiconductor thin films, and nanowires with the proximity induced superconductivity are proposed to provide an efficient platform for the realization of Majorana zero modes (MZMs)~\cite{Fu2008,Sau2010,Lutchyn10,Hughes2010,Oreg2010}. The latter are also experimentally realized in recent past~\cite{Mourik1003,Zareapour2012,Finck2013,He294,Liu18}. In such heterostructures, the MZMs are usually localized at two dimensional (2D) vortex cores or one dimensional (1D) edges where the topological superconducting gap in the bulk spectrum changes its sign. Very recently, the conventional bulk-boundary correspondence has been generalized in the context of higher-order topological insulators (HOTI) and higher-order topological superconductors (HOTSC)~\cite{PhysRevB.92.085126, benalcazar2017,benalcazarprb2017,Song2017,Langbehn2017,schindler2018,Khalaf2018,Geier2018,Franca2018,Zhu2018,Liu2018,Yan2018,wang2018higher,PhysRevB.98.165144,Ezawakagome,ezawa2019second,Roy2019,PhysRevB.99.045441,Trifunovic2019,Zhang2019,Volpez2019,YanPRB2019,Wu2020,Sumathi2020,jelena2020HOTSC,PhysRevB.101.220506,SongboPRR,SongboPRB,SongboPRR2,GhorashiPRL,GhorashiPRB,ZengPRL2020}. Precisely, an $n^{\rm th}$-order topological insulator or superconductor in $m$ dimensions hosts $d_c=(m-n)$-dimensional boundary modes ($n\le m$). For example, a three dimensional (3D) second (third) order topological  insulator (SOTI) hosts gapless modes on the hinges (corners), characterized  by $d_c=1~(0)$.  In particular, the SOTI phase has been experimentally realized in acoustic materials~\cite{XueAcousticKagome}, photonic crystals~\cite{PhotonicChen,PhotonicXie}, and electrical circuit~\cite{Imhof2018} setups.

Non-equilibrium aspects of topological phases have attracted a great deal of attention in the community as the driven topological systems exhibit non-trivial properties which are absent in the corresponding static phase~\cite{lindner2011floquet,Dora12,Rudner2013,Thakurathi2013,rechtsman2013photonic,maczewsky2017observation,Eckardt2017}. The Floquet machinery allows one to keep track of the time-dependent problem of periodically driven systems in a time-independent way with an effective Floquet Hamiltonian, defined in the frequency space~\cite{Shirley1965,Grossmann1991}. Therefore, the equilibrium notion of the topological invariant can be extended to Floquet topological phases where anomalous edge states or Floquet Majorana modes between two consecutive Floquet Brillouin zones appear~\cite{Rudner2013,oka2019,PhysRevB.90.205127}. Interestingly, Floquet engineering by suitably tuning appropriate perturbation can lead to Floquet HOTI  phases starting from a lower order or non-topological phases~\cite{Nag19,YangPRL2019,Seshadri2019,chaudhary2019phononinduced,Martin2019,Ghosh2020,Huang2020,PhysRevLett.124.057001,PhysRevB.101.085401,PhysRevResearch.1.032013,ZhangYang2020,ApoorvTiwari2020,YangPRR2020,Nag2020}. Therefore, a bunch of fundamentally important questions naturally arises (a) can the Floquet HOTSC phase be engineered by periodically driving the appropriate perturbation? (b) how does one topologically characterize the Floquet HOTSC phase?  In this paper, we intend to address these intriguing questions which have not been reported so far in the literature, to the best of our knowledge.

\begin{figure}
	\subfigure{\includegraphics[width=0.48\textwidth]{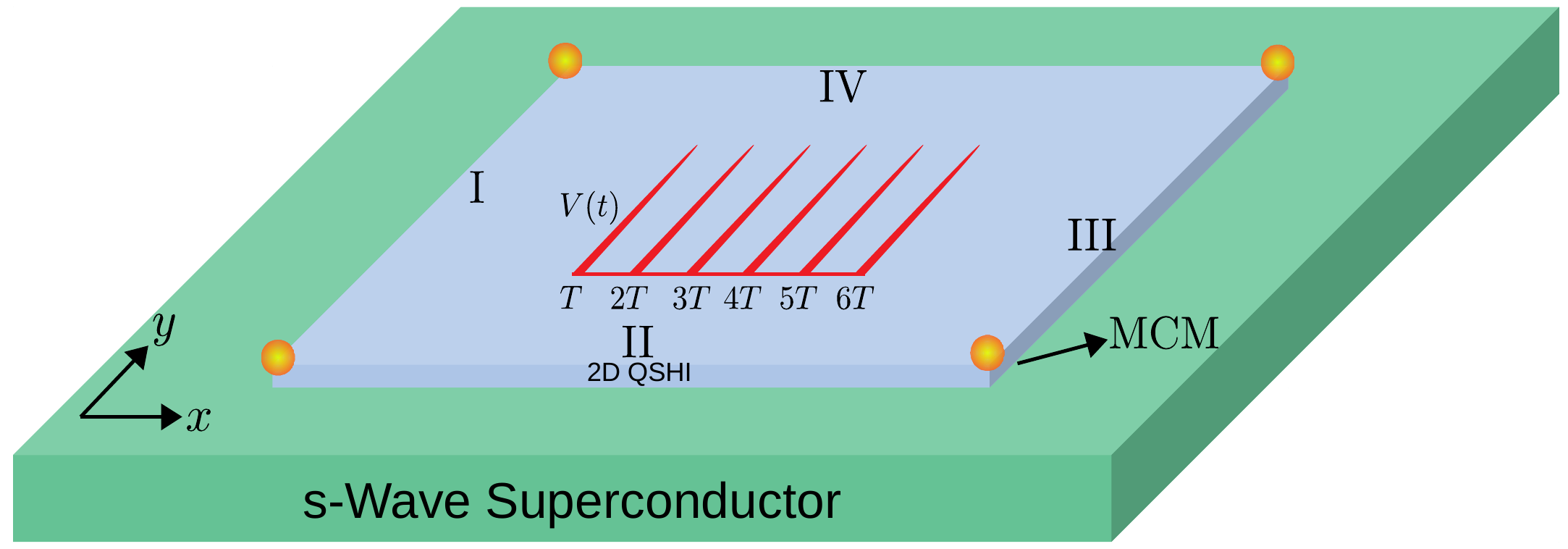}}
	\caption{(Color online) Schematic of our setup is illustrated in presence of periodic kick (red, grey) as an external drive. Here, a 2D QSHI (violet, light grey) is placed in close proximity to a bulk $s$-wave superconductor (green, grey).  MCMs are shown by circular dots (orange, light grey) and the four edges of the 2D QSHI are denoted by I, II, III, IV. 
}
	\label{model}
\end{figure}

In this article, we demonstrate a general mechanism of engineering the Floquet SOTSC phase by periodically kicking the QSHI Hamiltonian, proximitized by a $s$-wave superconductor. For the first case, we consider the periodic kicking in time-reversal symmetry (TRS) ${\mathcal T}$ breaking magnetic field to obtain the SOTSC phase hosting MCMs. In our second case, we introduce the kicking in four fold rotational symmetry $C_4$ and ${\mathcal T}$ breaking term so that the underlying edge states of QSHI phase become gapped and MCMs appear due to Jackiw-Rebbi index theorem~\cite{jackiw1976solitons}. Our dynamical model is schematically illustrated in Fig.~\ref{model} where the 2D QSHI/$s$-wave superconductor heterostructure is depicted with the general dynamical kicking protocol of perturbation $V(t)$. The MCMs appear between adjacent boundaries (edge I, II, III, IV).  We analytically derive the effective edge Hamiltonian for both the driven cases mentioned above to analyze the domain wall formation associated with the sign change of Dirac mass of the underlying low energy Hamiltonian. These SOTSC phases are appropriately characterized by both Floquet Wannier spectra (FWS) and Floquet quadrupolar moment (FQM).

The remainder of the paper is organized as follows. In Sec. \ref{sec:II}, we introduce our model Hamiltonian and the driving protocol. In Sec. \ref{sec:III}, we provide the effective Floquet Hamiltonian and illustrate the emergence of Floquet MCMs in the local density of states (LDOS) behavior. In Sec. \ref{sec:IV}, we characterize the higher order topological phase hosting MCMs by employing the appropriate topological invariants FQM and FWS that are extensively discussed in Appendix~\ref{A1}. In Sec. \ref{sec:V}, we resort to low energy edge theory to understand the analytical solutions of the zero energy MCMs. We provide the detailed analysis of that in Appendix~\ref{A2} and Appendix~\ref{A3}. In Sec. \ref{sec:VI}, we provide our alternative dynamical protocol to realize the MCMs and the necessary calculational details are properly supplemented in Appendix~\ref{A2} and  Appendix~\ref{A3}. Finally in Sec. \ref{sec:VII}, we summarize and conclude our results. 

\section{Model Hamiltonian and driving protocol}{\label{sec:II}}

We begin with the static Hamiltonian of a 2D QSHI placed in close proximity to a bulk $s$-wave superconductor~\cite{Wu2020},
\begin{eqnarray}
H_0= \vect{N}(k)\cdot {\bm \Gamma}\ ,
\label{ham1}
\end{eqnarray}
with $\vect{N}(k)=(N_1(k),N_2(k),N_3(k),N_4)$, $N_1(k)=2 \lambda_x \sin k_x$, $N_2(k)=2 \lambda_y \sin k_y$, $N_3(k)=\xi_k=(m_0-2t_x\cos k_x-2t_y\cos k_y)$, and $N_4=\Delta$. Here, $t_{x,y}$ and 
$\lambda_{x,y}$ represent the nearest-neighbor hopping and spin-orbit coupling respectively, $\Delta$ is the superconducting gap induced via the proximity effect, $m_0$ is the crystal-field splitting energy and $\mu$ is the chemical potential. Also, $\Gamma_1=\sigma_x s_z$, $\Gamma_2=\sigma_y \tau_z$, $\Gamma_3=\sigma_z \tau_z$, and  $\Gamma_4= s_y \tau_y$.
The three Pauli matrices ${\bm \sigma}$, ${\bm s}$ and ${\bm \tau}$ act on orbital $(a,b)$, spin $(\uparrow, \downarrow)$ and particle-hole degrees of freedom respectively. 
We work in the following basis, $C_k=\left(c_{k,a\uparrow}, c_{-k,a\uparrow}^\dagger, c_{k,a\downarrow}, c_{-k,a\downarrow}^\dagger, c_{k,b\uparrow}, c_{-k,b\uparrow}^\dagger, c_{k,b\downarrow}, c_{-k,b\downarrow}^\dagger\right)^T$. We consider chemical potential $\mu=0$ in order to obtain analytical results for the edge modes, otherwise, $\mu \tau_z $ can be added to the Hamiltonian (Eq.(\ref{ham1})). 

The Hamiltonian represented by Eq.(\ref{ham1}) preserves TRS ${\mathcal T}=i s_y K $ with $K$ being the complex conjugation.
If $\Delta=0$, the QSHI phase is observed when $[m_0^2 -(2 t_x + 2 t_y)^2][m_0^2-(2 t_x - 2 t_y)^2]<0$~\cite{FuKane2007}, hosting gapless propagating helical edge modes~\cite{bernevig2006quantum,konig2007quantum,hsieh2008topological}. 
When $\Delta \ne 0$, superconducting gap opens both in the bulk and helical edge spectrum and the system becomes a trivial BCS superconductor~\cite{Wu2020}. 
Interestingly, Hamiltonian (Eq.(\ref{ham1})) continues satisfying the unitary chiral symmetry ${\mathcal P}= \sigma_x s_y \tau_z$ and anti-unitary particle-hole symmetry ${\mathcal C}=\tau_x K$. 
These two symmetries turn out to be very important in determining the robustness of the SOTSC phase. 

We now introduce our driving protocol in the form of periodic kick as follows
\begin{eqnarray}
V(t)&=&h_x \Gamma_5 \sum_{r=1}^{\infty} \delta(t-rT) \ , 
\label{kick1}
\end{eqnarray}
and 
\begin{eqnarray}
V(t)  &=&\Lambda(k) \Gamma_6 \sum_{r=1}^{\infty} \delta(t-rT)\ ,
\label{kick2}
\end{eqnarray}
where, Eq.(\ref{kick1}) represents the TRS breaking driving protocol due to the in-plane Zeeman field $h_x$ applied along $x$-direction, $T$ is the period of the drive, $\Gamma_5=s_x\tau_z$ and 
$\Gamma_6=\sigma_x s_x \tau_z$. Here, the physical meaning of $\Gamma_{6}$ can be understood in terms of some hopping parameter that simultaneously flips both the orbital and spin.
The detailed outcome of $C_4$ and TRS symmetry breaking driving protocol (Eq.(\ref{kick2})) will be discussed in Sec~\ref{sec:VI}. In the static limit, $H^{I}_{\rm sta}=H_0 +h_x \Gamma_5$, 
is found to host MCMs in the SOTSC phase when $h_x > \Delta$~\cite{Wu2020}. The quasi-particle bandgap of the edges does not close along $k_y$ direction while the gap can be tuned along $k_x$ 
direction resulting in a topological phase transition and exponentially localized MCMs appear at zero energy protected by ${\mathcal P}$ and ${\mathcal C}$ symmetries.  Although, the bulk always remains gapped. 

\section{Effective Floquet Hamiltonian}{\label{sec:III}}
Following the periodic kick (see Eq.(\ref{kick1})), the Floquet operator reads  
\begin{eqnarray}
U(T)&=&{\widetilde{\mathcal {TO}}} \exp \left[-i\int_{0}^{T}dt\left(H_0+V(t)\right)\right] \nonumber \\
&=& \exp(-i H_0 T)~\exp(-i h_x \Gamma_5)\ .
\label{fo}
\end{eqnarray}
We can write $U(T)$ in a more compact form as
\begin{eqnarray}
U(T)&=&C_T\left(n_0-i n_5 \Gamma_5\right)-i S_T\sum_{j=1}^{4}\left(m_j\Gamma_j+p_j\Gamma_{j5}\right)\ ,
\end{eqnarray}
where, $C_T=\cos(\left|\vect{N}(k)\right| T)$, $S_T=\sin(\left|\vect{N}(k)\right| T)$, $n_0=\cos h_x$, $n_5=\sin h_x$, $m_j=\frac{N_j(k)\cos h_x}{\left|\vect{N}(k)\right|}$, $p_j=\frac{N_j(k)\sin h_x}{\left|\vect{N}(k)\right|}$ and $\Gamma_{j5}=\frac{1}{2i}\left[\Gamma_j,\Gamma_5\right]$ with $j=1,2,3,4$. The general form of the effective Hamiltonian thus found to be 
\begin{widetext}
\begin{eqnarray}
H^{I}_{\rm eff} =\frac{\epsilon_k}{\sin\epsilon_kT}\Bigg[\sin(\left|\vect{N}(k)\right| T) \cos h_x \sum_{j=1}^{4}n_j \Gamma_j 
+\cos(\left|\vect{N}(k)\right| T) \sin h_x \Gamma_5 +\sin(\left|\vect{N}(k)\right| T) \sin h_x \sum_{j=1}^{4} n_j \Gamma_{j5} \Bigg]\ ,
\end{eqnarray}
\end{widetext}

with $\epsilon_k=\frac{1}{T}\cos^{-1} \left[\cos(\left|\vect{N}(k)\right| T) \cos h_x\right]$, $n_j=\frac{N_j(k)}{\left|\vect{N}(k)\right|}$. In the high-frequency limit \ie~$T \rightarrow 0$ and $h_x \rightarrow 0$, neglecting the higher-order terms in $T$ and $h_x$, we find 
\begin{equation}
H^{I}_{\rm eff}\approx\sum_{j=1}^{4}N_j(k)\Gamma_j+\frac{h_x}{T} \Gamma_5+h_x\sum_{j=1}^{4}N_j(k)\Gamma_{j5}\ .
\label{heff1}
\end{equation}

\begin{figure}
	\centering
	\subfigure{\includegraphics[width=0.48\textwidth]{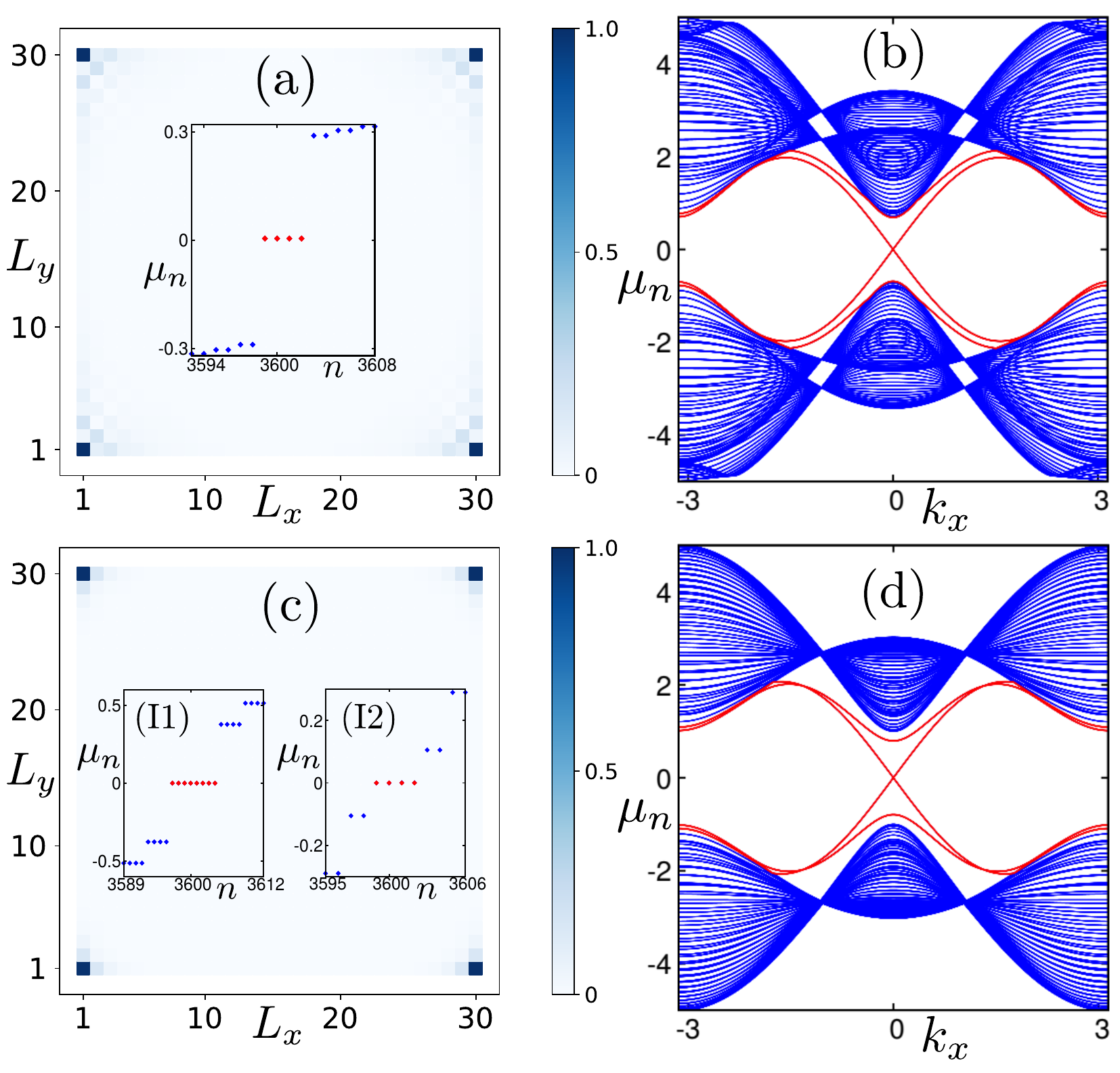}}
	\caption{(Color online)~(a) LDOS in finite geometry is demonstrated for driving protocol (Eq.(\ref{kick1})) and inset exhibits the eigenvalue spectrum for the same.  Here, $L_x=L_y=30, m_0=t_x=t_y=\lambda_x=\lambda_y=1.0$, $\Delta=0.4,h_x=0.4,T=0.628$. (b) The edge spectrum is shown for open boundary condition along $y$ direction for Eq.(\ref{heff1}). The edge gap closes when 
$\Delta=\frac{h_x}{T}$. (c) LDOS in finite geometry and eigenvalue spectrum (insets (I1) and (I2)) are depicted for driving protocol (Eq.(\ref{kick2})). We choose $\Lambda=0.3$ and the value of the 
other parameters remain the same as panel (a). For insets (I1) and (I2) we choose the parameter regime as $\lvert\frac{m\Lambda}{2t_xT}\rvert, \lvert\frac{m\Lambda}{2t_yT}\rvert > \Delta$ and 
$\lvert\frac{m\Lambda}{2t_xT}\rvert< \Delta < \lvert\frac{m\Lambda}{2t_yT}\rvert$ respectively.
(d) The edge spectrum is shown for open boundary condition along $y$ direction for the effective Hamiltonian (Eq.(\ref{heff2})). The edge gap closes when $\Delta=\lvert\frac{m\Lambda}{2t_xT}\rvert$. 
See text for discussion. 
	}
	\label{LDOS}
\end{figure}

Note that, in Eq.(\ref{heff1}), terms associated with $\Gamma_{j5}$ appears due to the drive 
and are absent in the static model; interestingly, only $\Gamma_{15}$ among $\Gamma_{j5}$ does not commute with $H^{I}_{\rm sta}$. 
As a result, $H^{I}_{\rm eff}$ loses the chiral symmetry generated by the unitary operator ${\mathcal P}$. Remarkably,  $H^{I}_{\rm eff}$ preserves the particle-hole symmetry which allows the generation of Floquet SOTSC phase with MCMs after the dynamical breaking of TRS. We now tie up our analytical finding by numerically diagonalizing the exact Floquet operator (Eq.(\ref{fo})) in the open boundary condition. One can obtain Floquet quasi-states $|\phi_n\rangle$ and quasi-energies $\mu_n$ from $U(T)$: $U(T)|\phi_n\rangle=\exp(-i \mu_n T)|\phi_n\rangle$. We present the local density of states (LDOS) associated with the zero (within numerical accuracy) quasi-energy Floquet quasi-states in Fig~\ref{LDOS}(a). These zero energy quasi-states correspond to the MCMs which are localized at the four corners 
of the system. 

\section{Topological characterization of MCMs}{\label{sec:IV}}

\begin{figure}
	\centering
	\subfigure{\includegraphics[width=0.48\textwidth]{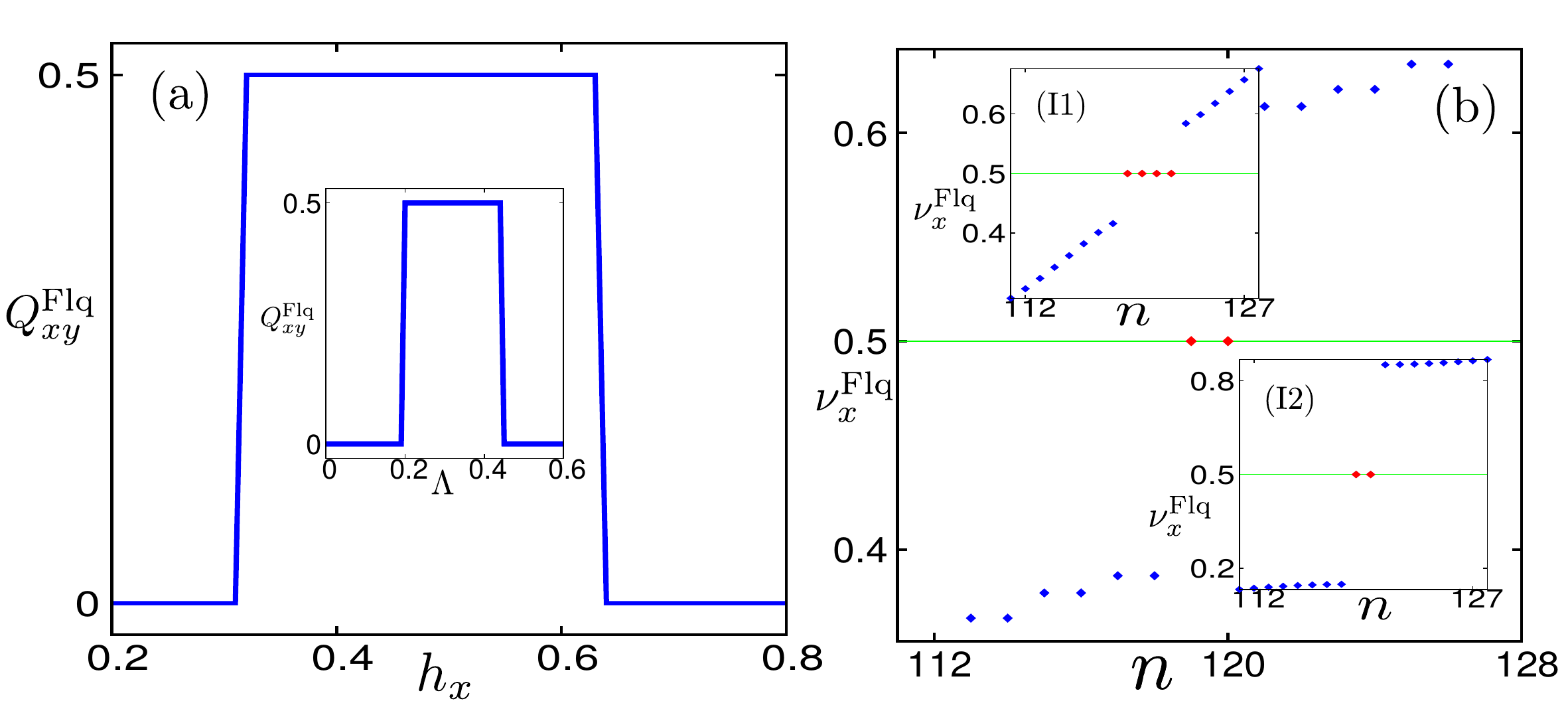}}
	\caption{(Color online) (a) Variation of FQM $Q^{\rm Flq}_{xy}$ is demonstrated as a function $h_x$ for driving protocol (Eq.(\ref{kick1})). In the inset, the same has been depicted with respect to 
$\Lambda$ for driving protocol (Eq.(\ref{kick2})). We choose $L_x=L_y=16, m_0=t_x=t_y=\lambda_x=\lambda_y=1.0$, $\Delta=0.4, T=0.628$. (b)~FWS $\nu^{\rm Flq}_x$ with respect to the state index is shown for driving protocol (Eq.(\ref{kick1})) and (Eq.(\ref{kick2}), insets (I1) and (I2)) respectively. For driving protocol (Eq.(\ref{kick1})), we choose $h_{x}>\Delta T$. Insets (I1) and (I2) have been depicted for the parameter regime $\lvert\frac{m\Lambda}{2t_xT}\rvert, \lvert\frac{m\Lambda}{2t_yT}\rvert > \Delta$ and $\lvert\frac{m\Lambda}{2t_xT}\rvert< \Delta < \lvert\frac{m\Lambda}{2t_yT}\rvert$ respectively for 
driving protocol (Eq.(\ref{kick2})). 
	}
	\label{invariant}
\end{figure}

To analyze the topological robustness of the Floquet MCMs in the SOTSC phase, we numerically compute the FQM (based on Floquet quasi-states 
obtained from numerical diagonalization of Eq.(\ref{fo})) $Q^{\rm Flq}_{xy}$ and the FWS $\nu^{\rm Flq}_x$ (based on $T\rightarrow 0$ effective Hamiltonian Eq.(\ref{heff1})) as shown in Fig.~\ref{invariant}(a) and (b) respectively. At the outset, we note that in the static limit, these invariants possess a quantized value of $0.5$. The FQM is defined through the Floquet many-body ground state $n_{F}$ composed by 
arranging the occupied quasi-states columnwise associated with the quasi-energy $-\omega/2 \le \mu_n \le 0$: $n_F=\sum_{n \in \mu_n \le 0} |\phi_n\rangle \langle \phi_n |$~\cite{Nag19,Ghosh2020}. 
We obtain $Q^{\rm Flq}_{xy}\equiv{\rm mod}(Q^{\rm Flq}_{xy},1)= 0.5$, this quantization (for a finite range of $h_{x}$ in protocol (Eq.(\ref{kick1}))) is depicted in Fig.~\ref{invariant}(a) and clearly suggests 
that the SOTSC phase, hosting MCMs, is topologically robust. Furthermore, we compute another invariant namely, the eigenvalue $\nu^{\rm Flq}_x$ of Floquet Wannier Hamiltonian 
${\mathcal H}^{\rm Flq}_{{\mathcal W}_x}$. The FWS $\nu^{\rm Flq}_x~(\nu^{\rm Flq}_y)$, demonstrated in Fig.~\ref{invariant}(b), exhibits two isolated eigenvalues at $0.5$ referring to the signature of MCMs in SOTSC phase for this protocol. We refer the readers to Appendix~\ref{A1} for the detailed calculation of FWS $\nu^{\rm Flq}_x~(\nu^{\rm Flq}_y)$.

\section{Low energy edge theory}{\label{sec:V}}

Here we proceed to derive the edge theory for the Floquet case starting from the effective Hamiltonian (Eq.(\ref{heff1})) in $T \rightarrow 0$ limit. 
The low energy effective Hamiltonian around $\Gamma=(0,0)$ point can be written as
\begin{eqnarray}\label{model1lowenergy}
H^{I}_{\rm eff, k}&\approx&(m+t_xk_x^2+t_yk_y^2)\Gamma_3+2\lambda_xk_x\Gamma_1+2\lambda_yk_y\Gamma_2\non \\
&& +\Delta \Gamma_4+\frac{h_x}{T} \Gamma_5+2\lambda_xh_xk_x\Gamma_{15}\ ,
\end{eqnarray}
where, $m=(m_0-2t_x-2t_y)$ and we assume $m<0$ in order to satisfy the topological condition \cite{FuKane2007}. We consider here the minimal model as $\Gamma_{15}$ is only incorporated among all the 
$\Gamma_{j5}$. As an representative example, for edge-I, we employ open (periodic) boundary condition along $x$ ($y$) direction. One can thus rewrite $H_{\rm eff, k}^{I}=H_I (-i \partial_x) + H_P(-i \partial_x,k_y, h_{x})$ neglecting $k_y^2$ term. Here, 
$H_I=(m-t_x \partial_x^2)\Gamma_3-2i\lambda_x\partial_x\Gamma_1$
and $H_P=2\lambda_yk_y\Gamma_2+\Delta \Gamma_4 +\frac{h_x}{T} \Gamma_5-2i\lambda_xh_x\partial_x\Gamma_{15}$.
Assuming $\Psi$ to be the zero energy eigenstate of $H_I$, we obtain (see Appendix~\ref{A2} for details)
\begin{eqnarray}
\Psi_{\alpha}=\left|\mathcal{N}_x\right|^2e^{-\mathcal{A}x}\sin \mathcal{B}x e^{ik_yy}\Phi_{\alpha}\ ,
\end{eqnarray}
where,
$\mathcal{A}=\frac{\lambda_x}{t_x}$, $\mathcal{B}=\sqrt{\left|\frac{m}{t_x}\right|-\mathcal{A}^2}$, $\left|\mathcal{N}_x\right|^2=\frac{4\mathcal{A}\left(\mathcal{A}^2+\mathcal{B}^2\right)}{\mathcal{B}^2}$ and $\Phi_{\alpha}$ is a 8-component spinor satisfying $\sigma_ys_z\tau_z\Phi_{\alpha}=-\Phi_{\alpha}$. We choose the following basis $\Phi_1=\ket{\sigma_y=+1}\otimes\ket{s_z=+1}\otimes\ket{\tau_z=-1},
\Phi_2=\ket{\sigma_y=-1}\otimes\ket{s_z=+1}\otimes\ket{\tau_z=+1},
\Phi_3=\ket{\sigma_y=-1}\otimes\ket{s_z=-1}\otimes\ket{\tau_z=-1},
\Phi_4=\ket{\sigma_y=+1}\otimes\ket{s_z=-1}\otimes\ket{\tau_z=+1}$
to cast the effective Hamiltonian for the edge-I as $H_i^{I,\rm Edge}=-2\lambda_yk_ys_z-\Delta s_y\tau_y$. Now the low energy effective Hamiltonian for the $l^{\rm th}$ edge is given as
(see Appendix~\ref{A2} for details)
\begin{equation}
\label{Model1edge}
H_{l}^{I,\rm Edge}=-iA_{l}s_z\partial_{l}-iB_{l}s_y\tau_z\partial_{l}-\Delta s_y\tau_y-h_{l}s_x\tau_z\ ,
\end{equation} 
with $A_{l}=\left\{-2\lambda_y,2\lambda_x,-2\lambda_y,2\lambda_x\right\}$, $B_{l}=\left\{0,-2\lambda_xh_x,0,-2\lambda_xh_x\right\}$ and $h_{l}=\left\{0,\frac{h_x}{T},0,\frac{h_x}{T}\right\}$. 
One can thus observe that the superconducting pairing gap has been induced in all the helical edge states irrespective of the Zeeman fields in that edge as $\{s_z,s_y\tau_y \}=0$.
  On the other hand, $h_x$ can only open up a Zeeman gap on two parallel edges (II and IV) without affecting two other perpendicular edges (I and III). We depict the quasi-energy spectrum of the semi-infinite slab geometry in Fig.~\ref{LDOS}(b) to manifest the gap closing at $\Delta=\frac{h_x}{T}$. 

In $\tau_x=\pm s_z$ subspace, the last term in Eq.(\ref{Model1edge}) can be written as $h_{l}s_x\tau_z\rightarrow \mp h_{l}s_y\tau_y$. 
We eventually obtain two decoupled diagonal blocks with Dirac masses $\Delta\pm \frac{h_x}{T} $ for edge II. While for edge I, the Dirac masses are of the same sign in these blocks. Therefore, for 
$h_x > \Delta T$, Dirac masses on edges I and II carry opposite sign leading to localized MCMs at the intersection of two perpendicular edges.
Interestingly, compared to the static case, the MCMs in Floquet SOTSC phase can be observed for a much smaller value of the in-plane magnetic field $h_x$ as $T\to 0$ in the high-frequency limit. 
One can find the wave-function for the zero-energy MCM, localized at the intersection between edge-I and II, as 
$\Psi_C \sim \exp \left[-\left(\Delta/2\lambda_y\right) y\right]\Phi_C $ and $\Psi_C \sim \exp \left[-\left(\lvert h_x-\Delta T\rvert/2\lambda_xT\right)x\right]\Phi_C$ for edges I and II, respectively, 
with $\Phi_C=\{1,-1,i,i\}^T$ (see Appendix~\ref{A3} for the derivation). The localization length of MCM can be different in different directions when $|h_x-\Delta T|/ 2 \lambda_xT \ne \Delta / 2 \lambda_y$ 
and it can be controlled along edge II by the period or frequency of the Floquet driving.

\section{Alternative dynamical protocol for realizing MCMs}{\label{sec:VI}}

Having established a route to obtain Floquet SOTSC starting from a static proximity induced QSHI, we here demonstrate another driving 
protocol to obtain the same by breaking the $C_4$ symmetry and TRS of the QSHI phase instead of an in-plane Zeeman field discussed before. We work with the dynamical protocol given in Eq.(\ref{kick2})
where, $\Lambda(k)=\Lambda (\cos k_x-\cos k_y)$. Upon adding this term with the static Hamiltonian (Eq.(\ref{ham1})), $H^{II}_{\rm sta}= H_0 + \Lambda(k) \Gamma_6$, one can find MCMs localized at zero energy and protected by unitary ${\mathcal P}$ and anti-unitary ${\mathcal C}$ symmetry~\cite{YanPRB2019}. Following our dynamical protocol, we numerically diagonalize the Floquet operator (Eq.(\ref{fo})) for this case and find strong corner localization of Majorana modes. In this case, eight Floquet MCMs appear at quasi-energy $\mu_n=0$ (within numerical accuracy) as shown in Fig.~\ref{LDOS}(c)-(I1). The MCMs obtained here are characteristically different from that of the originated by kicking the in-plane magnetic field where four zero energy MCMs are observed. For $h_x$ kicking (Eq.(\ref{kick1})), individual single-particle state residing at $\mu_n=0$ exhibits Majorana localization at a single corner. On the other hand, for $\Lambda(k)$ kicking (Eq.(\ref{kick2})), we find there exist at least two single-particle states sharing individual corner. 

For this protocol also, we derive the Floquet operator as follows,
$U(T)=C_T\left(n_0-i n_6 \Gamma_6\right)-i S_T\sum_{j=1}^{4}\left(m_j\Gamma_j+p_j\Gamma_{j6}\right)$
where, $C_T=\cos(\left|\vect{N}(k)\right| T)$, $S_T=\sin(\left|\vect{N}(k)\right| T)$, $n_0=\cos \Lambda(k)$, $n_6=\sin \Lambda(k)$, $m_j=\frac{N_j(k)\cos \Lambda(k)}{\left|\vect{N}(k)\right|}$, $p_j=\frac{N_j(k)\sin \Lambda(k)}{\left|\vect{N}(k)\right|}$ and $\Gamma_{j6}=\frac{1}{2i}\left[\Gamma_j,\Gamma_6\right]$. In the high-frequency limit $T \rightarrow 0$ and $\Lambda \rightarrow 0$, the effective Floquet Hamiltonian takes the form
\begin{equation}
H^{II}_{\rm eff}\approx\sum_{j=1}^{4}N_j(k)\Gamma_j+\frac{\Lambda(k)}{T} \Gamma_6+\Lambda(k)\sum_{j=1}^{4}N_j(k)\Gamma_{j6}\ .
\label{heff2}
\end{equation}

Similar to the effective Hamiltonian (Eq.(\ref{heff1})), here the drive induced terms are associated with $\Gamma_{j6}$ which essentially break the unitary symmetry ${\mathcal P}$. 
Importantly, the anti-unitary particle-hole symmetry generated by ${\mathcal C}$ assure the zero energy states to be localized at the corners. Following the same procedure described for the previous protocol, we obtain the low energy effective Hamiltonian of the edges in the edge-coordinate $l$ as (see Appendix~\ref{A2} for details)
\begin{equation}
H_{l}^{II,\rm Edge}=-iA_{l}s_z\partial_{l}+iB_{l}s_x\partial_{l}-\Delta s_y\tau_y+\Lambda_{l}s_y\ ,
\label{Model2edge}
\end{equation} 
with $A_{l}=\left\{-2\lambda_y,2\lambda_x,-2\lambda_y,2\lambda_x\right\}$, $B_{l}=\left\{\frac{m\lambda_y\Lambda}{t_x},\frac{m\lambda_x\Lambda}{t_y},\frac{m\lambda_y\Lambda}{t_x},\frac{m\lambda_x\Lambda}{t_y}\right\}$ and $\Lambda_{l}=\left\{-\frac{m\Lambda}{2t_xT},\frac{m\Lambda}{2t_yT},-\frac{m\Lambda}{2t_xT},\frac{m\Lambda}{2t_yT}\right\}$. 
For more insight, let us first consider $\Delta=0$. It is evident that $\Lambda_{l}$ changes sign at each corner and these lead to a domain wall formation of Dirac mass causing zero-energy 
Jackiw-Rebbi modes to appear at the corners~\cite{jackiw1976solitons,Nag19,Ghosh2020}. 
Due to the inclusion of the superconducting correlation, $H_{l}^{II, \rm Edge}$ can be decomposed into two independent parts as
\begin{eqnarray}
H^{II, \rm Edge}_l=H_{\tau_y=+1}\oplus H_{\tau_y=-1}\ ,
\end{eqnarray}
where,
\begin{eqnarray}
H_{\tau_y=+1}&=&-iA_{l}s_z\partial_{l}+iB_{l}s_x\partial_{l}+\left[-\Delta+\Lambda_{l}\right] s_y\non\\
H_{\tau_y=-1}&=&-iA_{l}s_z\partial_{l}+iB_{l}s_x\partial_{l}+\left[\Delta+\Lambda_{l}\right] s_y\ .
\label{ham2_edge}
\end{eqnarray}

The domain walls for both the sectors $\tau_y=\pm 1$ appear when $\lvert\frac{m\Lambda}{2t_xT}\rvert, \lvert\frac{m\Lambda}{2t_yT}\rvert > \Delta$. As a result, one finds two MCMs solutions 
(see Fig. \ref{LDOS}(c)-(I1)) per corner with the superposed wave-function $\Psi_C\sim \alpha \ e^{-\frac{M_I-\Delta}{2\lambda_y} y} \Phi_C^1 \ + \ \beta \ e^{-\frac{M_I+\Delta}{2\lambda_y} y} \Phi_C^2$ 
for edge-I. This phenomenon does not appear in the $h_x$-kick case where the domain wall for the Dirac mass appears only in one block of the edge Hamiltonian and the other block remains inactive (massive).
On the other hand, for $\lvert\frac{m\Lambda}{2t_xT}\rvert< \Delta <\lvert \frac{m\Lambda}{2t_yT}\rvert$, the domain walls exist in $\tau_y=+1 $ block, but not in $\tau_y=-1$ block. 
This results in a situation where only one MCM can present at each corner (see Fig.~\ref{LDOS}(c)-(I2)) with the wave-function $\Psi_C\sim e^{-\frac{M_I+\Delta}{2\lambda_y} y} \Phi_C^1$ for edge-I. 
See Appendix~\ref{A3} for the derivation of the MCMs wave-functions. The gap closing at $\Delta=\lvert\frac{m\Lambda}{2t_xT}\rvert$ has been illustrated in Fig.~\ref{LDOS}(d) based on a slab geometry calculation. 

We also calculate the FQM which appears to be $Q^{\rm Flq}_{xy}=0.0~(0.5)$ (see inset of Fig.~\ref{invariant}(a)) when there exist eight (four) MCMs at zero energy (see Figs.~\ref{LDOS}(c)-(I1)
(\ref{LDOS}(c)-(I2))). This is because two MCMs sharing each corner can fuse to a fermionic mode resulting in vanishing $Q^{\rm Flq}_{xy}$. On the other hand, one can find four (two) 
$\nu^{\rm Flq}_{x(y)}=0.5$ eigenvalues associated with Floquet Wannier Hamiltonian for the SOTSC phase with eight (four) zero energy MCMs as shown in Figs.~\ref{invariant}(b)-(I1) (\ref{invariant}(b)-(I2)). While for $h_x$ kick, there exist only two $\nu^{\rm Flq}_{x}=0.5$ eigenvalues associated with four zero energy MCMs. Thus one can infer that eight (four) MCMs represent the {\it{weak (strong)}} 
SOTSC phase. However, this subtlety cannot be distinguished from the feature of LDOS (see Figs.~\ref{LDOS}(a), (c)).

\section{Summary and Conclusions}{\label{sec:VII}}
To summarize, in this article, we demonstrate two dynamical protocols to generate Floquet SOTSC phase hosting MCMs.
In particular, a kick in the in-plane Zeeman field $h_{x}$, breaking the TRS, can lead to a {\it{strong}} SOTSC phase hosting only one MCM at each corner.  
In comparison, a $C_4$ and TRS breaking perturbation $\Lambda(k)$ can lead to eight (four) MCMs referring to {\it{weak (strong)}} SOTSC phase. We investigate the emergence of dynamical 
MCMs, localized at zero quasi-energy, by numerically diagonalizing the exact Floquet operator and analytically from the effective edge Hamiltonian. We show that these MCMs are protected by the anti-unitary particle-hole symmetry. We also characterize these phases by appropriate topological invariants such as FQM ($Q^{\rm Flq}_{xy}$) and FWS ($\nu^{\rm Flq}_{x}$ ($\nu^{\rm Flq}_{y}$)).

As far as experimental feasibility of our setup is concened, superconductivity in QSHI can be induced via proximity effect (\eg $\rm NbSe_{2}$)~\cite{ExperimentQSHI.SC1,ExperimentQSHI.SC2} with an induced gap $\Delta\sim 0.7~\rm meV$~\cite{ExperimentQSHI.SC2}. In recent times, experimental advancements on the pump-probe techniques~\cite{Wang453,maczewsky2017observation,Experiment2016} have enabled one to observe Floquet topological insulators~\cite{Wang453}. Therefore, we believe that the signature of MCMs 
may be possible to achieve via pump-probe based local scanning tunneling microscope (STM) measurements~\cite{Experiment.MZM.STM} for an in-plane magnetic field $h_{x}\sim 7-8~\rm T$.
On the other hand, our alternative dynamical protocol for generating SOTSC phases can in principle be realized in optical lattice platform where spin-orbit coupling, flipping the spin, 
is theoretically proposed~\cite{soc_theory1,soc_theory2,soc_theory3,soc_theory4,soc_theory5} and experimentally realized~\cite{lin2011spin,soc_exp1,huang2016experimental,soc_exp3,soc_exp2}.  Moreover, the synthetic spin-orbit coupling can in principle be realized in acoustic material~\cite{deng2020acoustic}. In recent times, Floquet driving has been experimentally demonstrated 
in various meta-materials such as, piezoelectric material~\cite{Experiment2014}, acoustic system~\cite{Experiment2016} and photonic systems~\cite{rechtsman2013photonic} etc. In particular, 
anomalous Floquet topological insulator (AFTI) has been experimentally demonstrated in acoustic systems~\cite{Experiment2016}. The model of AFTI is a two dimensional (2D) coupled metamaterial ring lattice. One can introduce a orbital like degree of freedom (${\bm \sigma}$) to the rings with two artificial atoms (ring resonators) $A$ and $B$~\cite{PhysRevLett.122.014302}. Also one can define a pseudo-spin (${\bm s}$) for acoustic waves based on wave circulation direction in each ring within the lattice. Thus, introducing the ring pseudo-spin degree of freedom and the proper inter-ring coupling, 
a synthetic spin-orbit interaction can be successfully induced which breaks the pseudo-spin conservation~\cite{deng2020acoustic}. Therefore, acoustic wave carrying a pseudo-spin in one lattice ring may tunnel into the adjacent coupled ring with the pseudo-spin flipped and in that process both orbital and spin can be effectively flipped. Hence, our alternative dynamical protocol (Eq.~(\ref{kick2})) is pertinent as far as the experiments on optical lattices and acoustic systems are concerned.



\acknowledgments{}
We acknowledge SAMKHYA: High-Performance Computing Facility at Institute of Physics, Bhubaneswar, for our numerical computation.
\appendix
{   
	\setcounter{figure}{0}
	\renewcommand{\thefigure}{A\arabic{figure}}
	\section{Floquet Wannier Spectra} \label{A1}
	In the semi-infinite geometry (considering periodic boundary condition (PBC) and open boundary condition (OBC)) along $x$ and $y$ direction respectively), we construct the Wilson loop operator~ \cite{benalcazarprb2017} ${\mathcal W}_x=F_{x,k_x + (N_x -1) \Delta k_x } \cdots F_{x,k_x + \Delta k_x } F_{x,k_x} $ with $ \left[F_{x,k_x}\right]_{mn}=\langle \phi_{n, k_x + \Delta k_x} | \phi_{m,k_x} \rangle$, where $\Delta k_x= 2\pi /N_x$ ($N_x$ being the number of discrete points considered inside the Brillouin zone (BZ) along $k_x$) and $|\phi_{m,k_x} \rangle $ is the $m^{\rm th}$ occupied Floquet 
	quasi-state. The latter can be obtained by diagonalizing the effective Floquet Hamiltonian in the high-frequecy limit. One can thus obtain the Wannier Hamiltonian, ${\mathcal H}^{\rm Flq}_{{\mathcal W}_x}= -i \ln {\mathcal W}_x$, whose eigenvalues $2\pi \nu^{\rm Flq}_x$ correspond to the Floquet Wannier spectra (FWS). Here, $\nu^{\rm Flq}_x \equiv {\rm mod} (\nu^{\rm Flq}_x,1)$ is the Wannier center. 
	One can similarly find $\nu^{\rm Flq}_y$. The feature of FWS characterizes the topological phase transition from trivial to higher-order topological superconductor (HOTSC) phase in our case. 
	In the Floquet HOTSC phase it acquires a quantized value 0.5 as shown in Fig.~\ref{FWShxLambda}. 
	
	\begin{figure}[h]
		\centering
		\subfigure{\includegraphics[width=0.48\textwidth]{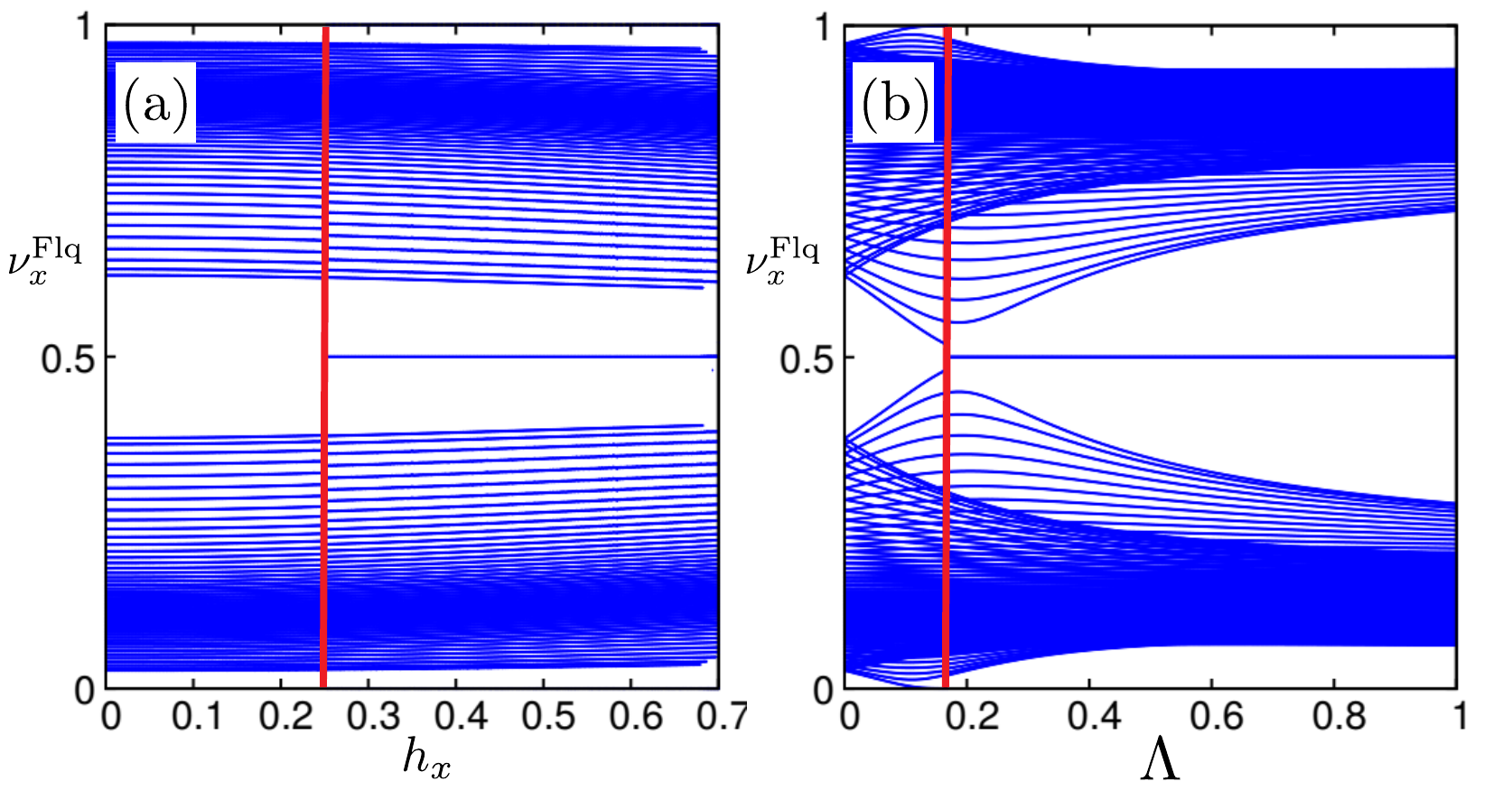}}
		\caption{(Color online) (a) FWS is shown as a function of the in-plane magnetic field $h_x$ for driving protocol 1 (Eq.~(\ref{kick1}) in the main text). (b) FWS is illustrated as a function 
			of the $C_{4}$ and $\mathcal{T}$ breaking mass term $\Lambda$ for driving protocol 2 (Eq.~(\ref{kick2}) in the main text). 
		}
		\label{FWShxLambda}
	\end{figure}
	
	\section{Low energy Edge theory} \label{A2}
	Here, we present the details of the low energy edge theory calculation for both our driving protocols.
	
	\subsection{Driving by in-plane magnetic field $h_{x}$}
	We begin by writting down the low energy effective Hamiltonian in the high-frequency limit (Eq.~(\ref{heff1}) in the main text) around $\Gamma=(0,0)$ point 
	\begin{eqnarray}\label{model1lowenergyA}
	H^{I}_{\rm eff, k}&\approx&(m+t_xk_x^2+t_yk_y^2)\Gamma_3+2\lambda_xk_x\Gamma_1+2\lambda_yk_y\Gamma_2\non \\
	&& +\Delta \Gamma_4+\frac{h_x}{T} \Gamma_5+2\lambda_xh_xk_x\Gamma_{15}\ ,
	\end{eqnarray}
	For edge-II, we consider PBC (OBC) along $x$ ($y$) direction. Hence, we replace $k_y$ by $-i\partial_y$ and rewrite $H_{\rm eff, k}^{I}=H_I (-i \partial_y) + H_P(-i \partial_y,k_x,h_x)$ neglecting $k_x^2$ term with 
	\begin{eqnarray}
	H_I&=&(m-t_y \partial_y^2)\Gamma_3-2i\lambda_y\partial_y\Gamma_2\ , \non \\
	H_P&=&2\lambda_xk_x\Gamma_1+\Delta \Gamma_4 +\frac{h_x}{T} \Gamma_5+2\lambda_xh_xk_x\Gamma_{15}\ , \quad\quad
	\end{eqnarray}
	Here, we choose $m<0$ to satisfy the Fu-Kane criteria~\cite{FuKane2007}. We solve $H_I$ exactly and treat $H_P$ as a perturbation. This approximation is valid when we assume the pairing amplitude, 
	$\Delta$ and the amplitude of the in-plane magnectic field, $h_x$ to be small~\cite{Wu2020}. We also consider any term multiplied by $h_x$ or $\Delta$ to be small.
	
	Assuming $\Psi$ to be the zero energy eigenstate of $H_I$, following the boundary condition $\Psi(0)=\Psi(\infty)=0$, we obtain
	\begin{equation}
	\Psi_{\alpha}=\left|\mathcal{N}_y\right|^2e^{-\tilde{\mathcal{A}}y}\sin \tilde{\mathcal{B}}y \ e^{ik_xx} \chi_{\alpha}\ ,
	\end{equation}
	where,
	$\tilde{\mathcal{A}}=\frac{\lambda_y}{t_y}$, $\tilde{\mathcal{B}}=\sqrt{\left|\frac{m}{t_y}\right|-\tilde{\mathcal{A}}^2}$, $\left|\mathcal{N}_y\right|^2=\frac{4\tilde{\mathcal{A}}\left(\tilde{\mathcal{A}}^2+\tilde{\mathcal{B}}^2\right)}{\tilde{\mathcal{B}}^2}$  and $\chi_{\alpha}$ is a 8-component spinor satisfying $\sigma_x\chi_{\alpha}=\chi_{\alpha}$. We work in the following basis for $\chi_\alpha$ as
	\begin{eqnarray}
	\chi_1&=&\ket{\sigma_x=+1}\otimes\ket{s_z=+1}\otimes\ket{\tau_z=-1},\non\\
	\chi_2&=&\ket{\sigma_x=+1}\otimes\ket{s_z=+1}\otimes\ket{\tau_z=+1},\non \\
	\chi_3&=&\ket{\sigma_x=+1}\otimes\ket{s_z=-1}\otimes\ket{\tau_z=-1},\non \\
	\chi_4&=&\ket{\sigma_x=+1}\otimes\ket{s_z=-1}\otimes\ket{\tau_z=+1}\ .
	\end{eqnarray}
	
	The matrix element of $H_P$ in this basis reads
	\begin{equation}
	H_{ii,\alpha\beta}^{I,\rm Edge}=\int_{0}^{\infty} dy \ \Psi_{\alpha}^\dagger(y) H_P(-i \partial_y,k_x,h_x)\Psi_{\beta}(y)\ ,
	\end{equation}
	Thus we obtain the effective Hamiltonian for the edge-II as
	\begin{equation}
	H_{ii}^{I,\rm Edge}=2 \lambda_x k_x s_z-2\lambda_xh_xk_x s_y\tau_z -\Delta s_y\tau_y -\frac{h_x}{T} s_x \tau_z\ ,
	\end{equation}
	Similarly, for edge-IV, we obtain the effective Hamiltonian as
	\begin{equation}
	H_{iv}^{I,\rm Edge}=2 \lambda_x k_x s_z-2\lambda_xh_xk_x s_y\tau_z -\Delta s_y\tau_y- \frac{h_x}{T} s_x \tau_z\ .
	\end{equation}
	
	For edge-III, we consider OBC (PBC) along $x$ ($y$) direction. One can thus rewrite $H_{\rm eff, k}^{I}=H_I (-i \partial_x) + H_P(-i \partial_x,k_y,h_x)$ by replacing $k_x \rightarrow -i\partial_x$ 
	and neglecting $k_y^2$ term with 
	\begin{eqnarray}
	H_I&=&(m-t_x \partial_x^2)\Gamma_3-2i\lambda_x\partial_x\Gamma_1\ , \non \\
	H_P&=&2\lambda_yk_y\Gamma_2+\Delta \Gamma_4 +\frac{h_x}{T} \Gamma_5-2i\lambda_xh_x\partial_x\Gamma_{15}\ . \quad\quad
	\end{eqnarray}
	
	Assuming $\Psi$ to be the zero energy eigenstate of $H_I$, following the boundary condition $\Psi(0)=\Psi(-\infty)=0$, and proceeding in a similar manner as before,
	we obtain
	\begin{equation}
	\Psi_{\alpha}=\left|\mathcal{N}_x\right|^2e^{\mathcal{A}x}\sin \mathcal{B}x \ e^{ik_yy} \Phi'_{\alpha}\ ,
	\end{equation}
	where, $\mathcal{A}=\frac{\lambda_x}{t_x}$, $\mathcal{B}=\sqrt{\left|\frac{m}{t_x}\right|-\mathcal{A}^2}$, $\left|\mathcal{N}_x\right|^2=\frac{4\mathcal{A}\left(\mathcal{A}^2+\mathcal{B}^2\right)}{\mathcal{B}^2}$ and $\Phi'_{\alpha}$ is a 8-component spinor satisfying $\sigma_ys_z\tau_z\Phi'_{\alpha}=\Phi'_{\alpha}$. We choose the following basis
	\begin{eqnarray}
	\Phi'_1&=&\ket{\sigma_y=-1}\otimes\ket{s_z=-1}\otimes\ket{\tau_z=+1}, \non \\
	\Phi'_2&=&\ket{\sigma_y=+1}\otimes\ket{s_z=-1}\otimes\ket{\tau_z=-1}, \non \\
	\Phi'_3&=&\ket{\sigma_y=+1}\otimes\ket{s_z=+1}\otimes\ket{\tau_z=+1}, \non \\
	\Phi'_4&=&\ket{\sigma_y=-1}\otimes\ket{s_z=+1}\otimes\ket{\tau_z=-1}\ .
	\end{eqnarray}
	
	The matrix element of $H_P$ in this basis can be written as
	\begin{equation}
	H_{iii,\alpha\beta}^{I,\rm Edge}=\int_{-\infty}^{0} dx \ \Psi_{\alpha}^\dagger(x) H_P(-i \partial_x,k_y,h_x)\Psi_{\beta}(x)\ ,
	\end{equation} 
	Thus we obtain the effective Hamiltonian for the edge-III as
	\begin{equation}
	H_{iii}^{I,\rm Edge}=-2 \lambda_y k_y s_z-\Delta s_y\tau_y\ .
	\end{equation}
	
	Therefore, the effective Hamiltonian for the four edges together can be written as
	\begin{eqnarray}
	H_{i}^{I,\rm Edge}&=&-2 \lambda_y k_y s_z-\Delta s_y\tau_y, \non \\
	H_{ii}^{I,\rm Edge}&=&2 \lambda_x k_x s_z-2\lambda_xh_xk_x s_y\tau_z -\Delta s_y\tau_y -\frac{h_x}{T} s_x \tau_z,\non \\
	H_{iii}^{I,\rm Edge}&=&-2 \lambda_y k_y s_z-\Delta s_y\tau_y,\non \\
	H_{iv}^{I,\rm Edge}&=&2 \lambda_x k_x s_z-2\lambda_xh_xk_x s_y\tau_z -\Delta s_y\tau_y- \frac{h_x}{T} s_x \tau_z\ . \non\\ 
	\end{eqnarray}
	
	\subsection{Driving by $C_4$ and $\mathcal{T}$ breaking mass term $\Lambda$}
	For this driving protocol also, we continue as before by writting down the low energy effective Hamiltonian in the high-frequency limit (Eq.~(\ref{heff2}) in the main text) around $\Gamma=(0,0)$ point as
	\small
	\begin{eqnarray}\label{model2lowenergy}
	H^{II}_{\rm eff,k}&=&(m+t_xk_x^2+t_yk_y^2) \Gamma_3+2\lambda_xk_x\Gamma_1+2\lambda_yk_y\Gamma_2+\Delta \Gamma_4 \non \\
	&&+\frac{\Lambda}{2T}\left(-k_x^2+k_y^2\right) \Gamma_6+\lambda_x\Lambda k_x\left(-k_x^2+k_y^2\right)\Gamma_{16}\non \\
	&&+\lambda_y\Lambda k_y\left(-k_x^2+k_y^2\right)\Gamma_{26} \non\\
	&&+\frac{\Lambda}{2}\left(m+t_xk_x^2+t_yk_y^2\right)\left(-k_x^2+k_y^2\right)\Gamma_{36}\ , 
	\end{eqnarray}
	\normalsize
	For edge-I, we consider OBC (PBC) along $x$ ($y$) direction and, as before we rewrite $H_{\rm eff, k}^{II}=H_I (-i \partial_x) + H_P(-i \partial_x,k_y,\Lambda)$. We replace $k_x\rightarrow -i\partial_x$ and neglect $k_y^2$ term. Thus we obtain 
	\begin{eqnarray}
	H_I&=&(m-t_x \partial_x^2)\Gamma_3-2i\lambda_x\partial_x \  \Gamma_1\ , \non \\
	H_P&=&2\lambda_yk_y\Gamma_2+\Delta \Gamma_4 +\frac{\Lambda}{2T}  \partial^2_x \ \Gamma_6-i\lambda_x\Lambda \ \ \partial_x^3\Gamma_{16}\non\\
	&&+\lambda_y \Lambda k_y \partial_x^2\Gamma_{26}+\frac{m\Lambda}{2}\partial_x^2\Gamma_{36}-\frac{\Lambda t_x}{2}\partial_x^4\Gamma_{36}\ , \qquad
	\end{eqnarray}
	Here, we consider the pairing amplitude $\Delta$ and the amplitude of the mass term $\Lambda$ to be small and treat them as small perturbation~\cite{Yan2018,YanPRB2019}. Assuming $\Psi$ to be the zero energy eigenstate of $H_I$ and following the boundary condition $\Psi(0)=\Psi(\infty)=0$, we obtain
	\begin{equation}
	\Psi_{\alpha}=\left|\mathcal{N}_x\right|^2e^{-\mathcal{A}x}\sin \mathcal{B}x \ e^{ik_yy} \Phi_{\alpha}\ ,
	\end{equation}
	where, $\mathcal{A}=\frac{\lambda_x}{t_x}$, $\mathcal{B}=\sqrt{\left|\frac{m}{t_x}\right|-\mathcal{A}^2}$, $\left|\mathcal{N}_x\right|^2=\frac{4\mathcal{A}\left(\mathcal{A}^2+\mathcal{B}^2\right)}{\mathcal{B}^2}$ and $\Phi_{\alpha}$ is a 8-component spinor satisfying $\sigma_ys_z\tau_z\Phi_{\alpha}=-\Phi_{\alpha}$. Our chosen basis reads
	\begin{eqnarray}
	\Phi_1&=&\ket{\sigma_y=+1}\otimes\ket{s_z=+1}\otimes\ket{\tau_z=-1}, \non \\
	\Phi_2&=&\ket{\sigma_y=-1}\otimes\ket{s_z=+1}\otimes\ket{\tau_z=+1}, \non \\
	\Phi_3&=&\ket{\sigma_y=-1}\otimes\ket{s_z=-1}\otimes\ket{\tau_z=-1}, \non \\
	\Phi_4&=&\ket{\sigma_y=+1}\otimes\ket{s_z=-1}\otimes\ket{\tau_z=+1}\ .
	\end{eqnarray}
	
	The matrix element of $H_P$ in this basis reads 
	\begin{equation}
	H_{i,\alpha\beta}^{II,\rm Edge}=\int_{0}^{\infty} dx \ \Psi_{\alpha}^\dagger(x) H_P(-i \partial_x,k_y,\Lambda)\Psi_{\beta}(x)\ ,
	\end{equation} 
	Thus we obtain the effective Hamiltonian for the edge-I as
	\begin{equation}
	H_{i}^{II,\rm Edge}=-2 \lambda_y k_y s_z-\frac{m\lambda_y \Lambda}{t_x}k_ys_x-\Delta s_y\tau_y-\frac{m\Lambda}{2t_xT}s_y\ ,
	\end{equation}
	Similarly, for edge-III, we obtain the effective Hamiltonian as
	\begin{equation}
	H_{iii}^{II,\rm Edge}=-2 \lambda_y k_y s_z-\frac{m\lambda_y \Lambda}{t_x}k_ys_x-\Delta s_y\tau_y-\frac{m\Lambda}{2t_xT}s_y\ .
	\end{equation}
	
	For edge-II, we employ OBC (PBC) along $y$ ($x$) direction. One can thus rewrite $H_{\rm eff, k}^{II}=H_I (-i \partial_y) + H_P(-i \partial_y,k_x,\Lambda)$ neglecting $k_x^2$ term 
	which yields 
	\begin{eqnarray}
	H_I&=&(m-t_y \partial_y^2)\Gamma_3-2i\lambda_y\partial_y\Gamma_2\ , \non \\
	H_P&=&2\lambda_xk_x\Gamma_1+\Delta \Gamma_4 -\frac{\Lambda}{2T}\partial^2_y \Gamma_6+i\lambda_y\Lambda\partial_y^3\Gamma_{26}\non\\
	&&-\lambda_x \Lambda k_x \partial_y^2\Gamma_{16}-\frac{m\Lambda}{2}\partial_y^2\Gamma_{36}+\frac{\Lambda t_x}{2}\partial_x^4\Gamma_{36}\ , \qquad
	\end{eqnarray}
	Assuming $\Psi$ to be the zero energy eigenstate of $H_I$, following the boundary condition $\Psi(0)=\Psi(\infty)=0$,
	we obtain
	\begin{equation}
	\Psi_{\alpha}=\left|\mathcal{N}_y\right|^2e^{-\tilde{\mathcal{A}}y}\sin \tilde{\mathcal{B}}y \ e^{ik_xx} \chi_{\alpha}\ ,
	\end{equation}
	where,
	$\tilde{\mathcal{A}}=\frac{\lambda_y}{t_y}$, $\tilde{\mathcal{B}}=\sqrt{\left|\frac{m}{t_y}\right|-\tilde{\mathcal{A}}^2}$, $\left|\mathcal{N}_y\right|^2=\frac{4\tilde{\mathcal{A}}\left(\tilde{\mathcal{A}}^2+\tilde{\mathcal{B}}^2\right)}{\tilde{\mathcal{B}}^2}$  and $\chi_{\alpha}$ is a 8-component spinor satisfying $\sigma_x\chi_{\alpha}=\chi_{\alpha}$. We choose the following basis
	\begin{eqnarray}
	\chi_1&=&\ket{\sigma_x=+1}\otimes\ket{s_z=+1}\otimes\ket{\tau_z=-1},\non\\
	\chi_2&=&\ket{\sigma_x=+1}\otimes\ket{s_z=+1}\otimes\ket{\tau_z=+1},\non \\
	\chi_3&=&\ket{\sigma_x=+1}\otimes\ket{s_z=-1}\otimes\ket{\tau_z=-1},\non \\
	\chi_4&=&\ket{\sigma_x=+1}\otimes\ket{s_z=-1}\otimes\ket{\tau_z=+1}\ .
	\end{eqnarray}
	
	The matrix element of $H_P$ in this basis can be written as
	\begin{equation}
	H_{ii,\alpha\beta}^{II,\rm Edge}=\int_{0}^{\infty} dy \ \Psi_{\alpha}^\dagger(y) H_P(-i \partial_y,k_x,\Lambda)\Psi_{\beta}(y)\ ,
	\end{equation}
	We obtain the effective Hamiltonian for the edge-II as
	\begin{equation}
	H_{ii}^{II,\rm Edge}=2 \lambda_x k_x s_z-\frac{m\lambda_x \Lambda}{t_y}k_xs_x-\Delta s_y\tau_y+\frac{m\Lambda}{2t_yT}s_y 
	\end{equation}
	Similarly, for edge-IV, we obtain the effective Hamiltonian as
	\begin{equation}
	H_{iv}^{II,\rm Edge}=2 \lambda_x k_x s_z-\frac{m\lambda_x \Lambda}{t_y}k_xs_x-\Delta s_y\tau_y+\frac{m\Lambda}{2t_yT}s_y\ .
	\end{equation}
	
	Therefore, the effective Hamiltonian for the four edges can be written as
	\begin{eqnarray}
	H_{i}^{II,\rm Edge}&=&-2 \lambda_y k_y s_z-\frac{m\lambda_y \Lambda}{t_x}k_ys_x-\Delta s_y\tau_y-\frac{m\Lambda}{2t_xT}s_y , \non \\
	H_{ii}^{II,\rm Edge}&=&2 \lambda_x k_x s_z-\frac{m\lambda_x \Lambda}{t_y}k_xs_x-\Delta s_y\tau_y+\frac{m\Lambda}{2t_yT}s_y ,\non \\
	H_{iii}^{II,\rm Edge}&=&-2 \lambda_y k_y s_z-\frac{m\lambda_y \Lambda}{t_x}k_ys_x-\Delta s_y\tau_y-\frac{m\Lambda}{2t_xT}s_y,\non \\
	H_{iv}^{II,\rm Edge}&=&2 \lambda_x k_x s_z-\frac{m\lambda_x \Lambda}{t_y}k_xs_x-\Delta s_y\tau_y+\frac{m\Lambda}{2t_yT}s_y\ .\non\\ 
	\end{eqnarray}
	\section{Majorana Corner Mode Solutions}{\label{A3}}
	Here, we provide the solutions for the zero energy MCMs for both the driving protocols. 
	\subsection{Driving by in-plane magnetic field $h_{x}$}
	To obtain the corner state solution (when $h_{x}>\Delta T$), in the intersection between edge-I and II, we solve the corresponding edge Hamiltonian for zero energy solution. 
	At edge-I, we assume a solution of the form 
	\begin{equation}
	\Psi_C \sim e^{-\lambda y}\Phi_C \ , 
	\end{equation}
	where, $\Phi_C$ is a four component spinor. The secular equation for $\Psi_C$ is given by
	\begin{equation}
	\det \left[ H_i^{I, \rm Edge}\right]=0\ ,
	\end{equation}
	We find four solutions for $\lambda$ as
	\begin{equation}
	\lambda=\left\{-\frac{\Delta}{2 \lambda_y},-\frac{\Delta}{2 \lambda_y},\frac{\Delta}{2 \lambda_y},\frac{\Delta}{2 \lambda_y}\right\}\ .
	\end{equation}
	
	As $\Psi_C$ must vanish at $y\rightarrow\infty$, therefore, we obtain two linearly independent solutions, $\Phi_C^1=\{1,1,-i,i\}^T$ and $\Phi_C^2=\{1,-1,i,i\}^T$. Thus, $\Psi_C$ can be expanded as
	\begin{equation}
	\Psi_C \sim \alpha \  e^{-\frac{\Delta}{2 \lambda_y} y}\Phi^1_C \ + \ \beta \  e^{-\frac{\Delta}{2 \lambda_y} y}\Phi^2_C\ ,
	\end{equation}
	Similarly, at edge-II, we obtain 
	\begin{equation}
	\Psi_C \sim \alpha' \  e^{-\frac{\lvert h_x-\Delta T \rvert }{2 \lambda_x T} x}\Phi^3_C \ + \ \beta' \   e^{-\frac{\lvert h_x+\Delta T \rvert }{2 \lambda_x T} x}\Phi^4_C \ ,
	\end{equation}
	where, $\Phi_C^3=\{1,-1,i,i\}^T$ and $\Phi_C^4=\{1,1,i,-i\}^T$. Considering the wavefuction, $\Psi_C$ to be continuous at the boundary \ie at $x=y=0$, we obtain $\alpha=\beta'=0$ and $\alpha'=\beta$. 
	Hence, the wavefunction for the Majorana corner mode becomes
	\begin{eqnarray}
	\Psi_C &\sim&  \  e^{-\frac{\Delta }{2 \lambda_y } y}\Phi^2_C \qquad \ {\rm :edge-I}\ , \non\\
	\Psi_C &\sim&  \  e^{-\frac{\lvert h_x-\Delta T \rvert }{2 \lambda_x T} x}\Phi^2_C \ \ {\rm :edge-II}\ . \non \\ 
	\end{eqnarray}
	with localization length $\left[\frac{\lvert h_x-\Delta T \rvert }{2 \lambda_x T}\right]^{-1}$ and $\left[\frac{\Delta}{2 \lambda_y }\right]^{-1}$ along $x$ and $y$ directions respectively. 
	Similarly, one can obtain the remaining zero energy corner mode solutions.
	\subsection{Driving by $C_4$ and $\mathcal{T}$ breaking mass term $\Lambda$}
	\subsubsection{Weak Phase}
	In this higher-order phase, $\lvert \frac{m\Lambda}{2t_xT} \rvert, \lvert \frac{m\Lambda}{2t_yT} \rvert > \Delta$. We proceed as before and obtain the following solutions at edge-I and II :
	\begin{eqnarray}
	\Psi_C&\sim& \alpha \ e^{-\frac{M_I-\Delta}{2\lambda_y} y} \Phi_C^1 \ + \ \beta \ e^{-\frac{M_I+\Delta}{2\lambda_y} y} \Phi_C^2 \quad  {\rm :edge-I}\ , \non\\
	\Psi_C&\sim& \alpha' \ e^{-\frac{M_{II}+\Delta}{2\lambda_x} x} \Phi_C^3+\  \beta' \ e^{-\frac{M_{II}-\Delta}{2\lambda_x} x} \Phi_C^4 \  {\rm :edge-II}\ , \non \\ 
	\end{eqnarray}
	with, $M_I=\lvert \frac{m\Lambda}{2t_xT} \rvert$, $M_{II}=\lvert \frac{m\Lambda}{2t_yT} \rvert$ and $\Phi_C^1=\{1,-i,1,-i\}^T$, $\Phi_C^2=\{1,i,1,i\}^T$, $\Phi_C^3=\{1,-i,1,-i\}^T$ and $\Phi_C^4=\{1,i,1,i\}^T$. Upon mathing the wavefunction at the boundary ($x=y=0$), we obtain $\alpha=\alpha'$ and $\beta=\beta'$. Hence, the final form of the zero energy solutions for the corner mode wavefunction reads
	\begin{eqnarray}
	\Psi_C&\sim& \alpha e^{-\frac{M_I-\Delta}{2\lambda_y} y} \Phi_C^1+ \beta e^{-\frac{M_I+\Delta}{2\lambda_y} y} \Phi_C^2 \ \ \  {\rm :edge-I}\ , \non\\
	\Psi_C&\sim& \alpha e^{-\frac{M_{II}+\Delta}{2\lambda_x} x} \Phi_C^1+ \beta e^{-\frac{M_{II}-\Delta}{2\lambda_x} x} \Phi_C^2 \  {\rm :edge-II}\ , \non \\ 
	\end{eqnarray}
	Thus, here two corner mode solutions exist for individual edge. 
	\subsubsection{Strong Phase}
	In this phase, we choose $M_I < \Delta < M_{II}$. We obtain the following solutions at edge-I and II :
	\begin{eqnarray}
	\Psi_C&\sim& \alpha \ e^{-\frac{M_I+\Delta}{2\lambda_y} y} \Phi_C^1 \ + \ \beta \ e^{-\frac{\Delta- M_I}{2\lambda_y} y} \Phi_C^2 \quad  {\rm :edge-I}\ , \non\\
	\Psi_C&\sim& \alpha' \ e^{-\frac{M_{II}-\Delta}{2\lambda_x} x} \Phi_C^3+\  \beta' \ e^{-\frac{M_{II}+\Delta}{2\lambda_x} x} \Phi_C^4 \  {\rm :edge-II}\ , \non \\ 
	\end{eqnarray}
	with $\Phi_C^1=\{1,i,1,i\}^T$, $\Phi_C^2=\{1,-i,-1,i\}^T$, $\Phi_C^3=\{1,i,1,i\}^T$ and $\Phi_C^4=\{1,-i,1,-i\}^T$. Therefore, mathing the wavefunction at the boundary, we obtain $\alpha=\alpha'$ 
	and $\beta=\beta'=0$. Hence, the final form of the solutions for the corner mode wavefunction becomes
	\begin{eqnarray}
	\Psi_C&\sim& e^{-\frac{M_I+\Delta}{2\lambda_y} y} \Phi_C^1 \ \ \  {\rm :edge-I}\ , \non\\
	\Psi_C&\sim& e^{-\frac{M_{II}-\Delta}{2\lambda_x} x} \Phi_C^1 \ \ {\rm :edge-II}\ . \non \\ 
	\end{eqnarray}
	In a similar fashion, one can obtain the zero energy solutions for the remaining MCMs in both these phases.
}

\bibliography{bibfile}{}

\end{document}